\documentclass{article}
\usepackage{authblk}
\usepackage[toc,page]{appendix}
\usepackage[left=1.5cm,right=1.5cm,top=2cm,bottom=2cm]{geometry}
\usepackage{amsmath,amsbsy,amsfonts,amssymb}
\usepackage[utf8x]{inputenc}
\usepackage{graphicx}
\usepackage{wasysym}
\usepackage{multicol}
\usepackage{fancyhdr}
\usepackage{hyperref}
\usepackage{color}
\usepackage{float}
\usepackage{enumitem}
\usepackage{subfloat}

\begin{document}

\title{Res-U2Net: Untrained Deep Learning for Phase Retrieval and Image Reconstruction}

\author[1]{Carlos~Osorio~Quero \footnote{ corresponding author: caoq@inaoep.mx}}
\author[2]{Daniel Leykam\footnote{daniel.leykam@gmail.com}}
\author[3]{Irving Rond\'on Ojeda \footnote{irving.rondon@gmail.com}}

\affil[1]{Computer Science Department, National Institute of Astrophysics, Optics and Electronics  (INAOE), Puebla 72810, Mexico}

\affil[2]{Centre for Quantum Technologies, National University of Singapore, 3 Science Drive 2,  Singapore 117543}

\affil[3]{School of Computational Sciences, Korea Institute for Advanced Study (KIAS), Seoul  02455, Republic of Korea}

%\affil[*]{ corresponding author: caoq@inaoep.mx}
%\affil[2]{daniel.leykam@gmail.com}
%\affil[3]{irving.rondon@gmail.com}%% email address is required; see note below about 
\date{}
\maketitle
\begin{abstract}
	Conventional deep learning-based image reconstruction methods require a large amount of training data which can be hard to obtain in practice.  Untrained deep learning methods overcome this limitation by training a network to invert a physical model of the image formation process. Here we present a novel untrained Res-U2Net model for phase retrieval. We use the extracted phase information to determine changes in an object's surface and generate a mesh representation of its 3D structure. We compare the performance of Res-U2Net phase retrieval against UNet and U2Net using images from the GDXRAY dataset.
\end{abstract}

%\begin{document}

	%%%%%%%%%%%%%%%%%%%%%%%%%%  body  %%%%%%%%%%%%%%%%%%%%%%%%%%
\section{Introduction}
	In recent times the field of computational imaging has witnessed significant advancements through the use of deep learning methods~\cite{optical_CI}. Deep learning has emerged as a promising approach for solving inverse problems encountered in computational imaging~\cite{optical_inv}. Groundbreaking studies have successfully demonstrated the effectiveness of deep learning for applications including optical tomography~\cite{optical_tomography}, 3D image reconstruction~\cite{Optical_3D,3D_PR_B}, phase retrieval~\cite{Optical_phase,Review_PR}, computational ghost imaging~\cite{Optical_ghost}, digital holography~\cite{Optical_holography,eHolo,Holographic}, imaging through scattering media~\cite{Optical_scattering}, fluorescence lifetime imaging under low-light conditions~\cite{Optical_fluorecence}, unwrapping~\cite{Optical_unwrapping,TIE}, and fringe analysis~\cite{Optical_finger}. 
	
	The deep learning-based artificial neural networks employed in computational imaging typically rely on a substantial collection of labeled data to optimize their weight and bias parameters through a training process~\cite{Training_DL}. This training enables the network to learn a universal function capable of mapping data from the object space to the image space. While traditional optimization methods struggle with this highly non-convex reconstruction problem, deep learning-based methods excel due to their nonlinear nature. Moreover, these methods can leverage statistical knowledge acquired from large datasets to infer solutions. Even though this reconstruction process is rapid in most cases~\cite{performance_DL}, the training procedure can be time-consuming, lasting several hours or even days, depending on the network architecture and the volume of data employed. Furthermore, obtaining a large and diverse training dataset, which is essential for the effectiveness of neural networks, is challenging. This is particularly relevant to phase retrieval problems where procuring an exhaustive collection of ground-truth images for training is frequently infeasible due to variability in the imaging apparatus. Limited training data then hampers the network's effectiveness and its capacity for generalization~\cite {untrained_DL}.
	
	Recent advances in imaging applications have demonstrated the immense potential of unsupervised learning techniques, specifically those utilizing untrained networks~\cite{DIP_DL,Y_Net,WPR}. By harnessing the inherent structure of neural networks without the need for training data, remarkable outcomes have been achieved. Two notable examples of untrained networks are the deep image prior~\cite{DIP_DL} and deep decoder~\cite{Non_Conv_DL}, which have effectively leveraged the network structure as a prior for image statistics, even without prior training. This approach involves employing a deep network with randomly initialized weights as an image generator to produce the recovered image. The network's weights are then repeatedly updated using a loss function that compares the generated image with the input data, such as a noisy image. This approach has demonstrated remarkable effectiveness in simulated image denoising~\cite{Non_Conv_DL}, deblurring~\cite{Untrained_Blur}, phase retrieval~\cite{Optical_phase,untrained_DL}, and super-resolution tasks~\cite{SR_Untrained}. 
	
	However, in many computational imaging problems, the acquired measurements do not directly resemble the reconstructed image. Instead, a forward model governs the relationship between the scene and the measurements, incorporating the underlying physics of the image formation problem. In the case of phase retrieval, this model constructs the phase of a sample using a known intensity distribution pair at the object and measurement planes. 
	
	There are a variety of amplitude-based phase retrieval algorithms with varying complexity, generality, and resolution~\cite{Model_PR_AMP}, including the Fourier-Born and Gerchberg-Saxton methods~\cite{GS_method,FB_method}. The former provides a closed-form solution to the phase retrieval problem under the assumption that the object only scatters the light weakly, while the latter is a more general but iterative method. Iterative phase retrieval algorithms often face convergence issues because the inverse problem is ill-posed: the phase profile that reproduces a given set of amplitude measurements is not unique, and minor changes in the measured data may result in significant uncertainties in the estimated solution~\cite{Ill_posed_P,ill_problem}. 
	
	One way to mitigate these challenges is to replace iterative methods with methods based on gradient descent. While the phase retrieval problem is typically a non-convex optimization problem, with many sub-optimal local minima, neural network parameterizations of the solution such as those based on UNet or other convolutional neural networks can be trained to obtain good solutions~\cite{UNet_DL,untrained_DL}.The UNet architecture consists of an encoder-decoder structure, where the encoder captures the high-level features from the input intensity data, and the decoder reconstructs the phase from these features~\cite{UNet_DL}. By incorporating physics-based constraints and priors into the network's design, a physics-enhanced deep neural network can learn to model the image formation process more accurately. This integration allows the network to exploit the known physics principles governing the imaging problem, thereby improving the rate of convergence compared to iterative phase retrieval methods such as the Gerchberg-Saxton algorithm and improving the quality and fidelity of the reconstructed images.
	
	In this study, we consider a method for the reconstruction of 2D and 3D images by combining diffraction models with convolutional neural networks and mesh estimation of the phase image~\cite{3D_mesh}. We consider three different neural networks for solving the inverse problem: UNet~\cite{UNet_DL} and U2Net~\cite{U2Net_DL}, which were previously used for unsupervised phase retrieval, as well as a new architecture called Res-U2Net. We show that the Res-U2Net-based architecture can achieve better performance compared to UNet and U2Net due to its ability to capture finer details of the image. Specifically, Res-U2Net incorporates additional layers in the form of downsampling and unsampling blocks, while preserving details with residual connections that skip some of the layers. To compare the performance of Res-U2Net against UNet and U2Net we use the GDXRAY~\cite{Gdxray_dataset} dataset of X-Ray images, which provides dimensional information. For the evaluation of the 2D image phase, we utilize standard metrics, namely No-Reference Image Quality Assessment (NR-IQA)~\cite{NR_IAQ}, Mean Squared Error (MSE)~\cite{3D_RMSE}, and skewness. These metrics provide insights into the performance, processing time, and quality of the reconstructed 3D images~\cite{3D_metric}.
	
	The outline of this article is as follows: Sec.~\ref{sec:Section_PM} briefly reviews prior works on phase retrieval, including the incorporation of deep learning-based techniques. Sec.~\ref{sec:resu2net} presents the architecture of our physics-informed Res-U2Net phase retrieval model, which we benchmark numerically in Sec.~\ref{sec:numerics}. We conclude with Sec.~\ref{sec:conclusion}.
	
	\section{Phase retrieval and deep learning}
	 \label{sec:Section_PM}

	The goal of the phase retrieval problem is to reconstruct an object's near field profile $\psi$ estimate from intensity measurements $I$, given information about the characteristics of the imaging system, represented by a (possibly nonlinear) operator $A$ that represents the imaging system, relating the object to the measured intensities~\cite{PR_mathematic,Fienup_PR},
	\begin{equation}
		I=\left | A \psi \right |^{2}. \label{eq:Phase}
	\end{equation}
	
	The absence of phase information makes the generic inverse problem challenging to solve~\cite{CH_PR_F}, as there exist infinitely many possible solutions that can yield the same set of measured intensities~\cite{Yiel_PR_F}. 
	\begin{equation}
		\tilde{\psi} = \mathrm{argmin}_{\psi} || \tilde{I} - |A\psi|^2 ||_2,
	\end{equation}
	
	Where, $\tilde{\psi}$ represents the estimated signal or image that we want to recover, $\psi$ represents the image whose phase information is to be retrieved, $\tilde{I}$ is the measured intensity, and $\mathrm{argmin}_{\psi}$ indicates that we are looking for the argument (in this case, $\tilde{\psi}$) that minimizes the following expression. Estimation of the signal $\tilde{\psi}$ given $I$ can be posed as a non-convex optimization problem~\cite{UNNphase}. The absence of phase information makes the generic inverse problem challenging to solve~\cite{CH_PR_F}, as there exist infinitely many possible solutions that can yield the same set of measured intensities~\cite{Yiel_PR_F}. 
	
	In many imaging setups, $A$ will simply be a matrix encoding the two-dimensional Fourier transform, which forms the basis for the Fourier-Born and Gerchberg-Saxton methods, among others. In this case, $I = I_z(x,y)$ is the intensity in the imaging plane $(x,y)$ at a distance $z$ from the object, which is related to the near field $\psi_0(x,y)$ via the transfer function $H_z$,
	\begin{equation} 
		I_z(x,y) = \left|\mathcal{F}^{-1}[ e^{-i kz \sqrt{1 - \lambda^2 (k_x^2 + k_y^2)}} \mathcal{F} [ \psi_0(x,y)] \right|^2 = |H_z \psi_0(x,y)|^2, \label{eq:diffraction}
	\end{equation}
	where $k = 2\pi/\lambda$, $\lambda$ is the imaging wavelength, and $\mathcal{F}$ is the two-dimensional Fourier transform. In the case of uniform illumination of a pure phase object, the near field can be written as $\psi_0(x,y) = I_0 e^{i \theta(x,y)}$ and the aim is to determine the phase profile $\theta(x,y)$ using $I_z(x,y)$.
	
	In the case of iterative phase retrieval methods, such as the Gerchberg-Saxton algorithm, one iteratively updates the estimated image $\tilde{\psi} \in \mathbb{C}^{N \times N}$ in the spatial and Fourier domains~\cite{GS_method}. However, these kinds of iteration procedures have several drawbacks. They tend to stagnate and exhibit slow convergence, often requiring more than 1000 iterations to reach a solution. In addition, they are highly sensitive to the initial conditions.  
	
	Alternatively, if the object is known to scatter the probe light only weakly (Born approximation), the near field can be written as $\psi_0(x,y) \approx I_0(1 +i \theta(x,y))$ and it follows that
	\begin{equation} 
		I_z(x,y) = I_0 (1 + 2 \mathrm{Re}[ H_z\theta(x,y)]), \label{eq:Born}
	\end{equation}
	allowing the near-field to be directly reconstructed from the Fourier transform of the far field intensity. For more information about Fourier phase retrieval algorithms, we recommend interested readers to consult Refs.~\cite{GS_method,FB_method,optical_tomography}.
	
	To address the limitations of conventional phase retrieval methods, newer gradient-based algorithms such as the Writinger Flow (WF) phase retrieval have been introduced~\cite{WF_PR}, which solve the phase retrieval problem~\eqref{eq:Phase} using gradient descent~\cite{AMP_PR},
	\begin{equation}
		\tilde{\psi}^{j+1}=\tilde{\psi}^{j}-\mu^{j+1} \nabla f(\tilde{\psi}^{j})\label{eq:PR_gradient},    
	\end{equation}
	where $\nabla f(\tilde{\psi}^{j})$ represents the first-order gradient of the loss function and $\mu^{j+1}$ denotes the step size at the current iteration $j$. WF offers theoretical guarantees for convergence to the globally minimal solution. Empirically, a set of 4 to 8 intensity iterations is usually enough for convergence to the globally optimal solution. Unfortunately, WF often fails to converge to a satisfactory outcome when provided with only a single intensity measurement, limiting its effectiveness in this case unless one can impose prior constraints on the form of the object.

	Recently extensive research has been conducted on phase retrieval using deep learning-based approaches. These approaches offer faster, non-iterative inferences compared to other more time-consuming optimization-based algorithms~\cite{Model_PR_AMP}. Most of these methods can reconstruct the phase using a single Fourier intensity measurement without any additional constraints. This seemingly ill-posed problem can have a unique solution (with only minor ambiguities) for the original complex-valued signal if the Fourier intensity measurement is oversampled by at least twice in each dimension~\cite{Fourier_PR,Model_PR_AMP}.  In general, deep learning-based unsupervised phase retrival methods can be categorized based on whether or not they incorporate the underlying physics into their networks~\cite{DL_PR_Methods}.
	
	The first category of deep learning-based methods involves using a feedforward network to directly estimate target images from a Fourier intensity measurement~\cite{DeepPhaseCut}. For instance, Ref.~\cite{Oversampled_Fourier} proposed a two-branch convolutional neural network (CNN) to reconstruct the magnitude and phase of 3D crystal image from an oversampled Fourier intensity measurement. While this approach can demonstrate reasonably good performance for simple images with limited details, its efficacy for complex images remains unknown. Other image reconstruction methods based on conditional generative adversarial networks~\cite{GAN_PR} and multiple multilayer perceptrons~\cite{ResNet_FR} tailored for Fourier phase retrieval tasks can similarly capture simple features adequately but give relatively large errors in capturing finer details~\cite{Lensless_FR}.
	
	The second category of deep learning-based phase retrieval methods aims to enhance the quality of reconstructed images by effectively leveraging the underlying physics into their models~\cite{FR_DL_2}. One such approach incorporates physics information through a learnable spectral initialization~\cite{LSpetral} followed by a double-branch UNet for reconstruction. However, this approach requires an additional masking scheme to impose constraints on the measurement, leading to rather noisy reconstructed images, even for simple images~\cite{Book_FR_DL}. Another proposed method involves the use of multi-layer perceptrons of different sizes in a cascaded network. In this approach, the intensity measurement is applied to each multi-layer perceptron to assist in training and inference~\cite{FR_Ite_DL}. Despite its merits, this approach struggles to reconstruct fine image details and necessitates a large network size due to the utilization of multiple multi-layer perceptrons~\cite{Book_convolutional}.
	
	To enhance the performance of the neural network model, it is beneficial to integrate physics-informed principles into a deep learning framework. This involves training the neural network to effectively learn the inverse mapping from the observed intensity data $I_z(x,y)$ to the corresponding near field phase profile $\theta(x,y)$ using a forward diffraction model~\cite{UNet_DL,Unet_PR_F1,Unet_PR_F2,Unet_PR_F3}. The estimated phase profile is then used as input into the diffraction model to obtain an estimate for the far field intensity; the difference between this estimate and the measured intensity serves as a loss function used to train the network and improve the phase estimate. UNet, a popular architecture in the field of image processing and computer vision, can learn the inverse of the diffraction operator~\cite{Unet_PR,Unet_PR_F1,untrained_DL,Unet_PR_F4}. 
	
	By integrating a UNet architecture within the phase retrieval process, we can harness the powerful representational capabilities of deep neural networks while retaining the convergence characteristics typical of traditional gradient descent-based phase retrieval algorithms. This UNet-based approach excels in capturing the complex interactions between intensity and phase data, rooted in the underlying physics of the problem. Additionally, the process involves a progressive refinement aspect, distinct from iterative methods in numerical analysis. This refinement, occurring over multiple cycles, allows for gradual improvement of the phase estimate. Such an approach is particularly beneficial in mitigating the challenges posed by initial conditions and noise in the intensity data, thereby enhancing the accuracy of phase retrieval~\cite{Unet_PR_F5}.

	%\subsection{Ill-Posed Problem of Phase Retrieval} 
	
	%In the domain of image reconstruction and related inverse problems, the issue of ill-posedness arises when even minor changes in measured data result in significant uncertainties in the estimated solution, as highlighted by previous research~\cite{Ill_posed_P,ill_problem}. Such scenarios manifest when critical information is missing, as exemplified in cases of image reconstruction where phase details are absent. The absence of phase information introduces an inherent ill-posed nature to the problem, as there exist countless potential solutions that can match the given amplitude measurements. Traditional methodologies and iterative algorithms face substantial challenges in achieving a definitive and accurate solution, especially when confronted with this ill-posed nature, further compounded by measurement noise.To mitigate these challenges, a novel approach has been proposed, involving the utilization of UNet-based techniques~\cite{UNet_DL}. UNet represents a deep learning architecture capable of learning intricate mappings between input and output data, thereby enhancing phase estimation and convergence for algorithms like GS and FR. The integration of classical algorithms with UNet presents a promising avenue for tackling the complexities posed by ill-posed problems, ultimately bolstering the robustness of image reconstruction.
	
\section{Phase retrieval using Res-U2Net}
	\label{sec:resu2net}
	
\begin{figure}
		\centering
		\includegraphics[width=\columnwidth]{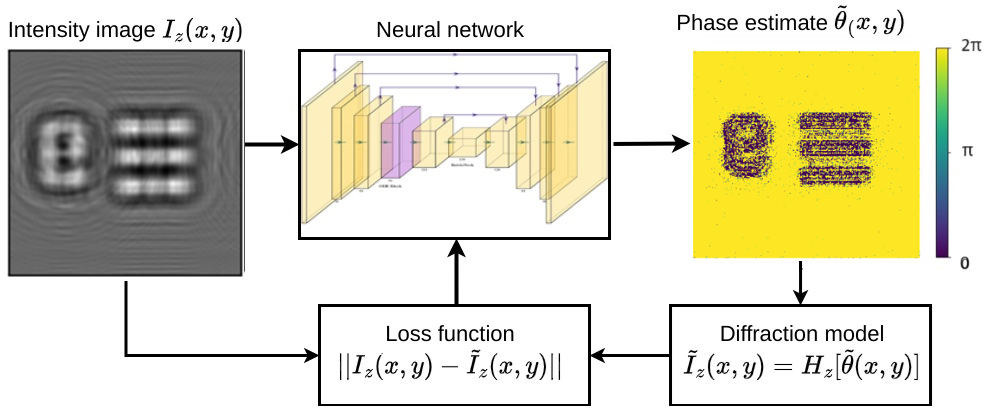}
		\caption{Schematic of the phase retrieval process. An intensity image $I_z(x,y)$ is fed into a neural network, returning an estimate of the near-field phase $\tilde{\theta}(x,y)$. Diffraction model $H_z$ converts the estimated near field phase to an estimated far field intensity profile $\tilde{I_z(x,y)}$. The mean square error (MSE) between $I_z(x,y)$ and $I_z(x,y)^{*}$ serves as a loss function for optimizing the parameters of the neural network.}
		\label{fig:untrained}
	\end{figure}

	The untrained phase retrieval process (see Fig.~\ref{fig:untrained}) involves applying a Fourier-based forward model~\cite{Model_PR_AMP}, which is defined in Section~\ref{Section_PM}, to evaluate the input image and obtain the image plane intensity $I_z(x,y)$. This diffraction model is then used as input for an untrained neural network that estimates the near field phase $\tilde{\theta}(x,y)$. We evaluate three different neural networks: UNet, U2Net, and Res-U2Net. The neural network is trained by comparing the intensity profile $\tilde{I}_z(x,y)$ obtained from the diffraction model $A$, specified by ~\eqref{eq:diffraction} or~\eqref{eq:Born}, using the estimated phase $\tilde{\theta}(x,y)$, and comparing against the measured far-field intensity $I_z(x,y)$. The neural network is trained by minimizing the mean square error (MSE) between the measured and estimated intensities, $||I_z(x,y) - \tilde{I}_z(x,y)||$, following the approach introduced in Ref.~\cite{Optical_phase}. This minimization of the cost function via gradient descent allows for the gradual refinement of the estimated phase until a desired accuracy, as measured by the difference between subsequent iterations, is obtained.
	
	\subsection{Res-U2Net Structure} 
	
\begin{figure*}[!t]
	\centering
	\includegraphics[width=7in]{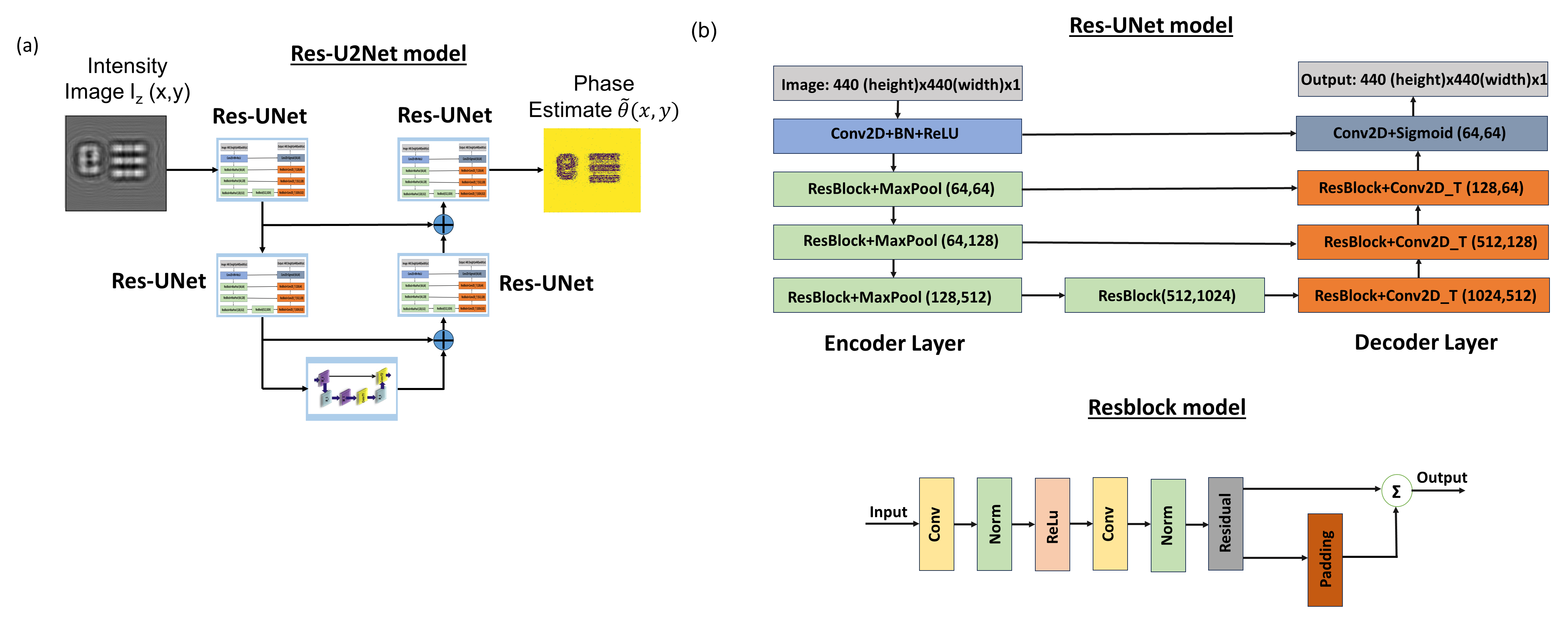}
	\caption{Res-U2Net architecture: (a) U2Net model of configuration based on sequence multi-scale, that integrates res-model in the network, (b) Res-UNet model, the encoder extracts features using convolutional layers (Conv2D) with batch normalization, ReLU activation (ResBlock), and spatial resolution reduction via max pooling (MaxPooling2D). This is followed by a decoder assigning phases to the features by upsampling using transpose convolutions (Conv2DTranspose) with skip connections. Residual connections link the encoder and decoder layers to improve the training performance. Finally, a $1\times440\times440$ convolutional layer generates the segmentation mask, resulting in the network output. }
	\label{fig:Res_U2NET}
\end{figure*}

	The Res-U2Net architecture, as shown in Fig.~\ref{fig:Res_U2NET}, is a sophisticated evolution of the UNet and Res-UNet models, particularly designed for image segmentation tasks~\cite{UNet_DL}. This architecture innovatively incorporates residual connections at various stages, enhancing information exchange and gradient flow during training. Structurally similar to UNet, it includes both downsampling and upsampling processes. The downsampling path consists of convolutional layers with batch normalization and ReLU activation, followed by max-pooling layers. Conversely, the upsampling path employs transposed convolutional layers to enlarge the feature maps. A key advancement of Res-U2Net over UNet is the amalgamation of a series of encoder/decoder modules that combine the Res-Unet model (refer to Fig.~\ref{fig:Res_U2NET}) with a series of stacked U-Nets. This configuration, featuring layers for both downsampling and upsampling, promotes more efficient feature transmission and mitigates the issue of vanishing gradients~\cite{Unet_vanish}.
	
	Res-U2Net has shown exceptional capabilities in various image segmentation tasks, surpassing the original UNet's performance. While its design is primarily for image segmentation, this does not directly imply its effectiveness in phase retrieval tasks, since phase retrieval often involves complex patterns rather than areas of near-uniform phase values, challenging the assumption that segmentation-focused methods would automatically excel here. Nevertheless, the U-Net architecture has been successfully applied to the phase retrieval ~\cite{Optical_phase}. 
	However, we aim to explore new possible architectures using familiar ones.

Res-U2Net performs the following standard sequence of operations.
(See Fig.~\ref{fig:Res_U2NET}a):
\begin{enumerate}
	\item Res-UNet: The architecture applies a series of Res-UNet blocks in parallel, allowing the network to process and extract features at different scales or levels of abstraction.
	\item Combination: The features extracted by the parallel Res-UNet blocks are combined to produce a richer set of features.
\end{enumerate}

In Fig.~\ref{fig:Res_U2NET}b), it is shown how Res-UNet carries out the encoder and decoder operations:
\begin{enumerate}
	%  \item Reshaping Input Tensor: The input tensor 'input' is reshaped to match the desired input dimensions. Here, 'batchsize' represents the number of images in a batch, and 'W' and 'H' represent the width and height of the input image.
	\item Feature Extraction (Encoder layer): This part of the model uses convolutional layers, batch normalization, and ReLU activation functions to process the input image. The convolutional layers are designed to extract features by applying filters that capture spatial hierarchies in the image. As the input passes through these layers, it is transformed into a set of feature maps that represent different aspects of the input. The batch normalization helps in stabilizing the learning process by normalizing the input of each layer, and the ReLU (Rectified Linear Unit) activation function introduces non-linearity, allowing the model to learn complex patterns.
	\item Spatial Resolution Reduction: Max pooling is used after convolutional blocks to reduce the spatial dimensions of the feature maps. This operation helps in making the representation smaller and more manageable, and it also introduces some level of invariance to small translations in the input image.
	\item Upsampling (Decoder layer): The decoder part of the network uses transpose convolutions (also known as up-convolutions or deconvolutions) to increase the spatial dimensions of the feature maps. This process is essential for tasks like image segmentation, where the goal is to produce an output image that is the same size as the input image.
	\item Skip Connections: These connections are used to combine feature maps from the downsampling path with those from the upsampling path. By doing so, the network can leverage both high-level semantic information and low-level spatial information, which is crucial for accurately reconstructing the output.
	\item Residual Connections: After concatenating feature maps from different layers, additional convolutions are applied, and their output is added element-wise to the input of the concatenation. This creates a residual block (Resblock), which helps mitigate the vanishing gradient problem and allows for deeper networks by promoting more effective backpropagation of gradients.
	\item Upsampling and Concatenation: This step repeats the process of upsampling and combining feature maps from different layers of the network. It ensures that as the spatial dimensions are restored to match the input size, the network progressively refines the details of the output.
	
	\item Segmentation Mask Generation: Finally, we introduced a convolutional layer with a sigmoid activation function to generate the final segmentation mask, serving as the output of the neural network. This step converts the input vector into a matrix, containing the estimated phase information. The resulting matrix matches the dimensions of the input image, specifically a size of $1\times440\times440$.
\end{enumerate}

\subsection{3D phase reconstruction}

\begin{figure}[h!]
	\centering \includegraphics[width=\columnwidth]{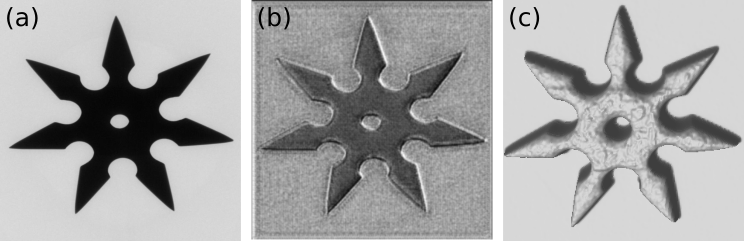}%
	\caption{3D Phase Retrieval: (a) 2D Ray-X test image, (b) 2D phase retrieval estimate, and (c) resulting 3D mesh.}
	\label{fig:3D_phase}
\end{figure}

We utilized the Unified Shape-From-Shading Model (USFSM) to perform the 3D reconstruction of the estimated image obtained through phase retrieval. The USFSM approach constructs three-dimensional representations by analyzing the spatial intensity variations present in the two-dimensional recovered image~\cite{eikonal}. To extract depth information from the phase retrieval image, which corresponds to the surface points of the scene, we employed the fast-sweeping method. This method employs the Lax–Friedrichs Hamiltonian technique~\cite{Lax_3D} to solve for the surface, utilizing an iterative sweeping strategy based on the fast sweeping scheme described in Ref.~\cite{3D_mesh}. An example is shown in Fig.~\ref{fig:3D_phase}.

\section{Results}
\label{sec:numerics}

\begin{figure}
	\centering
	\includegraphics[width=\columnwidth]{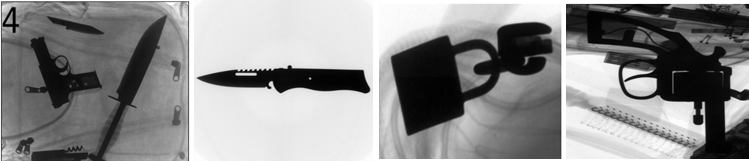}
	\caption{Examples from the GDXRAY dataset of $440 \times 440$ pixel images.}\label{fig:Test_image}
\end{figure}

We carried out numerical calculations to compare the performance of phase retrieval using Res-U2Net against the UNet and U2Net networks, considering Fourier and Fourier-Born diffraction models and four $440 \times 440$ pixel images from the GDXRAY dataset, shown in Fig.~\ref{fig:Test_image}.  In all cases, we limit the maximum number of training iterations to 1000. Training was performed using the Keras framework in Python~\cite{UNET_keras}, in the algorithm and defined a stopping criterion with a network error tolerance of $10^{-4}$. The calculations were executed using an NVIDIA GTX 1080 graphics processing unit (GPU).

The performance of the trained models was evaluated based on the quality of the reconstructed 2D images and 3D meshes. The configurations of the neural network and diffraction models were analyzed to determine the most efficient approach in terms of image, processing time, and mesh reconstruction quality.

\subsection{2D Phase Retrieval}

\begin{figure}
	\centering
	\includegraphics[width=\columnwidth]{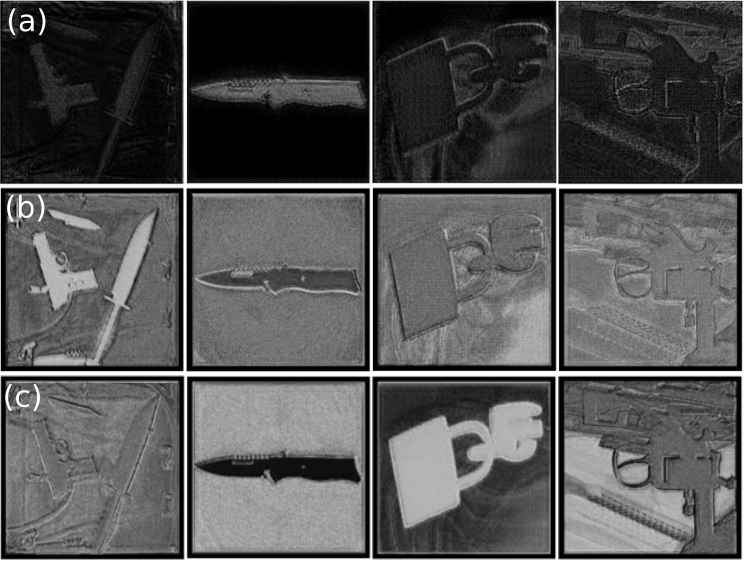}
	\caption{2D Fourier phase retrieval using (a) UNet, (b) U2Net, and (c) Res-U2Net.}\label{fig:figure_GS}
\end{figure}

\begin{figure}
	\centering
	\includegraphics[width=\columnwidth]{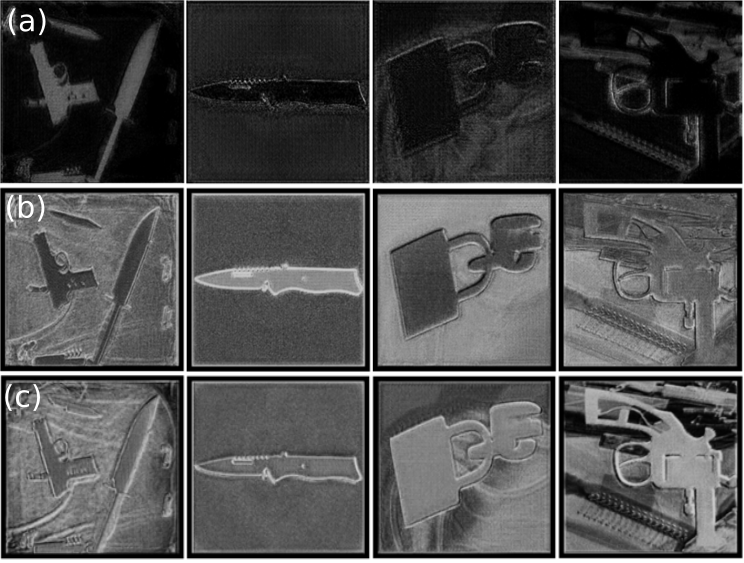}
	\caption{2D Fourier-Born phase retrieval using (a) UNet, (b) U2Net, and (c) Res-U2Net.}\label{fig:figure_FB}
\end{figure}

Figs.~\ref{fig:figure_GS} and~\ref{fig:figure_FB} present the estimated phase profiles obtained from the different neural networks using the Fourier and Fourier-Born diffraction models, respectively. The image processing times ranged from 0.5 to 5 seconds for images of differing complexity. The networks differ in the detail and contrast of the phase images produced.

To quantify the performance of the phase reconstruction we used NR-IAQ, considering first the BRISQUE (Blind/Referenceless Image Spatial Quality Evaluator) method~\cite{Brisque_NR_IAQ}. This method assesses the statistical properties of an image to estimate its quality, with lower scores indicating better image quality and higher scores indicating poorer quality. The BRISQUE scores presented in Table~\ref{tab:integrated1a} show that Res-U2Net consistently outperforms UNet and U2Net for both the generic Fourier imaging and the special case of weak phase contrast (Fourier-Born approximation). Res-U2Net consistently achieves the lowest BRISQUE scores, indicative of its superior ability to produce images with the highest perceived quality.

\begin{table}
	\centering
	\caption{Performance for 2D phase retrieval using UNet, U2Net, and Res-U2Net quantified using Blind/Referenceless Image Spatial Quality Evaluator (lower is better)~\cite{Brisque_NR_IAQ}.}
	\begin{tabular}{|l|l|l|l|}
		\hline
		Method & UNet & U2Net & Res-U2Net \\
		\hline
		Fourier & 16.37 & 11.67 & 9.63 \\
		\hline
		Fourier-Born & 15.17 & 9.04 & 8.12  \\
		\hline
	\end{tabular}
	\label{tab:integrated1a}
\end{table}

Next, we consider NIQE (Natural Image Quality Evaluator)~\cite{NIQE_NR_IAQ}, which assesses factors such as texture, sharpness, and entropy. Higher values of NIQE indicate images with lower perceptual quality, while lower values suggest higher quality. A similar trend to BRISQUE is visible in the NIQE scores presented in Table~\ref{tab:integrated1b}. Once again, Res-U2Net achieves the lowest scores across all image types, signifying superior image quality and enhanced contrast levels, as can be seen in Figs.~\ref{fig:figure_GS} and~\ref{fig:figure_FB}.

\begin{table}
	\centering
	\caption{Performance for 2D phase retrieval using UNet, U2Net, and Res-U2Net quantified using  Natural Image Quality Evaluator (lower is better)~\cite{NIQE_NR_IAQ}.}
	\begin{tabular}{|l|l|l|l|}
		\hline
		Method & UNet & U2Net & Res-U2Net \\
		\hline
		Fourier & 3.23 & 2.89 & 1.69 \\
		\hline
		Fourier-Born & 2.62 & 2.53 & 1.41 \\
		\hline
	\end{tabular}
	\label{tab:integrated1b}
\end{table}

\subsection{3D Retrieval Phase}

\begin{figure}
	\centering
	\includegraphics[width=\columnwidth]{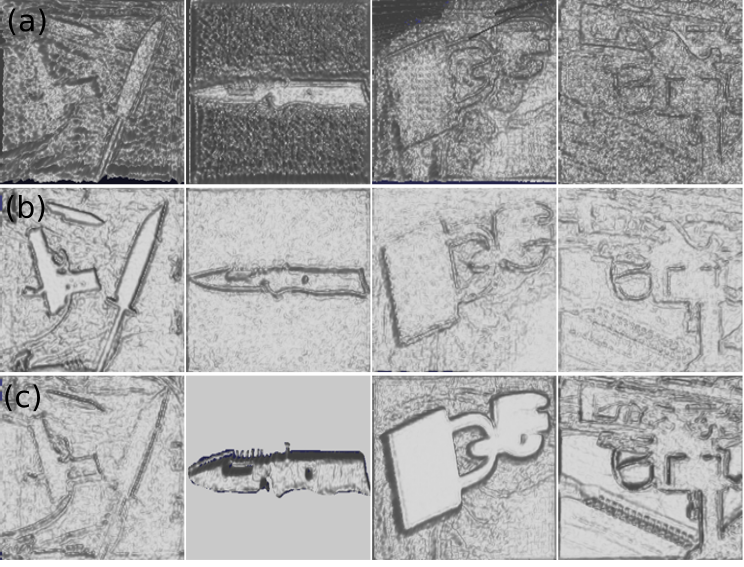}
	\caption{3D Fourier phase retrieval images obtained using (a) UNet, (b) U2Net, and (c) Res-U2Net.}\label{fig:figure_3DGS}
\end{figure}

\begin{figure}
	\centering
	\includegraphics[width=\columnwidth]{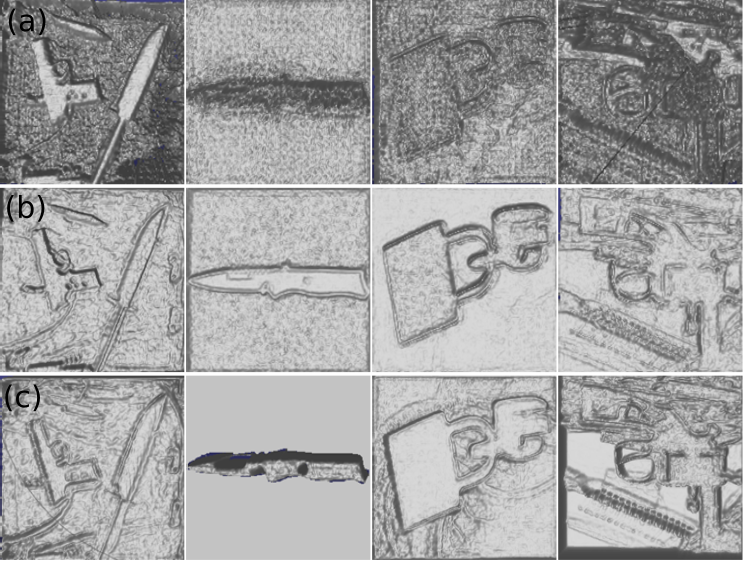}
	\caption{3D Fourier-Born phase retrieval images using (a) UNet, (b) U2Net, and (c) Res-U2Net.}\label{fig:figure_3DFB}
\end{figure}

Next we applied the Shape-From-Shading Model (USFSM) to reconstruct 3D images from the 2D phase profiles obtained from the Fourier and Fourier-Born diffraction models. The resulting images are presented in Figs.~\ref{fig:figure_3DGS} and~\ref{fig:figure_3DFB}.

To quantify the performance of the 3D image reconstruction, we calculated the Mean Squared Error (MSE) and Skewness ~\cite{3D_RMSE}, which defines the symmetry of the 3D shapes. A Skewness value near 0 indicates the best mesh and a value close to 1 indicates a completely degenerate mesh~\cite{3D_FOM} between the 3D mesh of the normalized test images (Fig.~\ref{fig:Test_image}) and the 3D mesh obtained from the phase reconstruction. The results can be found in Tables~\ref{tab:integrated2a} and ~\ref{tab:integrated2b}.

\begin{table}
	\centering
	\caption{Performance for 3D image reconstruction from the UNet, U2Net, and Res-U2Net phase images quantified by the Mean Squared Error (lower is better).}
	\begin{tabular}{|l|l|l|l|}
		\hline
		Method & UNet & U2Net & Res-U2Net \\
		\hline
		Fourier & 0.150 & 0.059 & 0.056  \\
		\hline
		Fourier-Born & 0.270 & 0.062& 0.053 \\
		\hline
	\end{tabular}
	\label{tab:integrated2a}
\end{table}

\begin{table}
	\centering
	\caption{Performance for 3D image reconstruction from the UNet, U2Net, and Res-U2Net phase images quantified by the Skewness (lower is better).}
	\begin{tabular}{|l|l|l|l|}
		\hline
		Method & UNet & U2Net & Res-U2Net \\
		\hline
		Fourier & 0.545 & 0.1460 & 0.1140 \\
		\hline
		Fourier-Born & 1.145 & 0.0100 & 0.0049 \\
		\hline
	\end{tabular}
	\label{tab:integrated2b}
\end{table}

Inspecting the MSE values, UNet has a somewhat worse performance compared to the other two neural networks for both diffraction models. Moving on to the Skewness values, again UNet has the worst performance, indicating lower spatial resolution in the reconstructed 3D image, limiting its ability to discern surface details of the tested objects, as is visible in Figs.~\ref{fig:figure_3DGS} and~\ref{fig:figure_3DFB}. Meanwhile, U2Net and Res-U2Net exhibit significantly better Skewness values for the Fourier-Born diffraction model. We also note that methods incorporating physics-based forward diffraction models tend to yield lower Skewness values compared to other neural network models.

\section{Conclusions}
\label{sec:conclusion}

In this study, we examined the use of physics-informed deep learning techniques for phase retrieval, considering the specific example of phase retrieval from X-ray images. Our main objective is to assess the efficacy of these methods in 2D and 3D imaging. We conducted a thorough analysis of three neural networks - UNet, U2Net, and Res-U2Net - to determine their suitability for unsupervised Fourier phase retrieval in X-ray imaging. Our findings reveal significant improvements in both 2D and 3D reconstructions, with processing times ranging from 0.5 to 5 seconds. Notably, we observed enhanced background details and improved Skewness scores in 3D meshes generated from GDXRAY test images. Res-U2Net, in particular, exhibited promising potential as a robust method for producing high-quality 2D Fourier phase retrieval images. On the other hand, UNet may require further optimizations to match the effectiveness of its counterparts. This investigation highlights the synergy between neural network models and physics-based forward models, offering an effective approach for phase retrieval tasks in 3D mesh normalized test images.

Future research should focus on exploring additional evaluation metrics and refining these models to optimize their performance in specific imaging applications. For example, the integration of Generative Adversarial Networks (GANs) may be used to increase the network's robustness against noise and decrease the number of artifacts incorporated during the image phase estimation~\cite{GAN_DL}. Res-U2Net can be adapted for phase retrieval of images in other spectral bands, and it can also be applied to other imaging problems found in different fields, such as biomedical imaging.

%\begin{backmatter}

\section*{Acknowledgments}
	C. O. Q. acknowledges financial support from CONACYT Mexico 251992. D. L. acknowledges support from the National Research Foundation, Singapore and A*STAR under its CQT Bridging Grant.
	I.R. is supported by the Basic Science Research Program through the National Research Foundation of Korea (NRF) funded by the Ministry of Science and ICT [NRF-2017R1E1A1A01077717].
	
\section*{Disclosures}
The authors declare no conflicts of interest.
	
\section*{Data Availability }
	The data and code underlying the results presented in this paper are not publicly available at this time but may be obtained from the authors upon reasonable request.
	
%\end{backmatter}

%%%%%%%%%% If using BibTeX:
\bibliographystyle{unsrt} 
\bibliography{sample}

\begin{thebibliography}{10}

\bibitem{optical_CI}
Danni Zhang and Zhongwei Tan.
\newblock A review of optical neural networks.
\newblock {\em Applied Sciences}, 12(11), 2022.

\bibitem{optical_inv}
Shuo Liu, Xiuguo Chen, Tianjuan Yang, Chunfu Guo, Jiahao Zhang, Jianyuan Ma,
  Chao Chen, Cai Wang, Chuanwei Zhang, and Shiyuan Liu.
\newblock Machine learning aided solution to the inverse problem in optical
  scatterometry.
\newblock {\em Measurement}, 191:110811, 2022.

\bibitem{optical_tomography}
Anna Burvall, Ulf Lundstr\"{o}m, Per A.~C. Takman, Daniel~H. Larsson, and
  Hans~M. Hertz.
\newblock Phase retrieval in {X}-ray phase-contrast imaging suitable for
  tomography.
\newblock {\em Opt. Express}, 19(11):10359--10376, May 2011.

\bibitem{Optical_3D}
Xian-Feng Han, Hamid Laga, and Mohammed Bennamoun.
\newblock Image-based {3D} object reconstruction: State-of-the-art and trends
  in the deep learning era.
\newblock {\em IEEE Transactions on Pattern Analysis and Machine Intelligence},
  43(5):1578--1604, 2021.

\bibitem{3D_PR_B}
Lei Tian and Laura Waller.
\newblock {3D} intensity and phase imaging from light field measurements in an
  {LED} array microscope.
\newblock {\em Optica}, 2(2):104--111, Feb 2015.

\bibitem{Optical_phase}
Fei Wang, Yaoming Bian, Haichao Wang, Meng Lyu, Giancarlo Pedrini, Wolfgang
  Osten, George Barbastathis, and Guohai Situ.
\newblock Phase imaging with an untrained neural network.
\newblock {\em Light: Science {\&} Applications}, 9(1):77, May 2020.

\bibitem{Review_PR}
Yoav Shechtman, Yonina~C. Eldar, Oren Cohen, Henry~Nicholas Chapman, Jianwei
  Miao, and Mordechai Segev.
\newblock Phase retrieval with application to optical imaging: A contemporary
  overview.
\newblock {\em IEEE Signal Processing Magazine}, 32(3):87--109, 2015.

\bibitem{Optical_ghost}
Tomoyoshi Shimobaba, Yutaka Endo, Takashi Nishitsuji, Takayuki Takahashi, Yuki
  Nagahama, Satoki Hasegawa, Marie Sano, Ryuji Hirayama, Takashi Kakue, Atsushi
  Shiraki, and Tomoyoshi Ito.
\newblock Computational ghost imaging using deep learning.
\newblock {\em Optics Communications}, 413:147--151, 2018.

\bibitem{Optical_holography}
Tianjiao Zeng, Yanmin Zhu, and Edmund~Y. Lam.
\newblock Deep learning for digital holography: a review.
\newblock {\em Opt. Express}, 29(24):40572--40593, Nov 2021.

\bibitem{eHolo}
Hao Wang, Meng Lyu, and Guohai Situ.
\newblock {eHoloNet}: a learning-based end-to-end approach for in-line digital
  holographic reconstruction.
\newblock {\em Opt. Express}, 26(18):22603--22614, Sep 2018.

\bibitem{Holographic}
Zhenbo Ren, Zhimin Xu, and Edmund Y.~M. Lam.
\newblock {End-to-end deep learning framework for digital holographic
  reconstruction}.
\newblock {\em Advanced Photonics}, 1(1):016004, 2019.

\bibitem{Optical_scattering}
Ryoichi Horisaki, Ryosuke Takagi, and Jun Tanida.
\newblock Learning-based imaging through scattering media.
\newblock {\em Opt. Express}, 24(13):13738--13743, Jun 2016.

\bibitem{Optical_fluorecence}
Jason~T. Smith, Nathan Un, Ruoyang Yao, Nattawut Sinsuebphon, Alena
  Rudkouskaya, Joseph Mazurkiewicz, Margarida Barroso, Pingkun Yan, and Xavier
  Intes.
\newblock Fluorescent lifetime imaging improved via deep learning.
\newblock In {\em Biophotonics Congress: Optics in the Life Sciences Congress
  2019}, page NM3C.4. Optica Publishing Group, 2019.

\bibitem{Optical_unwrapping}
Kaiqiang Wang, Ying Li, Qian Kemao, Jianglei Di, and Jianlin Zhao.
\newblock One-step robust deep learning phase unwrapping.
\newblock {\em Opt. Express}, 27(10):15100--15115, May 2019.

\bibitem{TIE}
Kaiqiang Wang, Jianglei Di, Ying Li, Zhenbo Ren, Qian Kemao, and Jianlin Zhao.
\newblock Transport of intensity equation from a single intensity image via
  deep learning.
\newblock {\em Optics and Lasers in Engineering}, 134:106233, 2020.

\bibitem{Optical_finger}
Shijie Feng, Qian Chen, Guohua Gu, Tianyang Tao, Liang Zhang, Yan Hu, Wei Yin,
  and Chao Zuo.
\newblock {Fringe pattern analysis using deep learning}.
\newblock {\em Advanced Photonics}, 1(2):025001, 2019.

\bibitem{Training_DL}
George Barbastathis, Aydogan Ozcan, and Guohai Situ.
\newblock On the use of deep learning for computational imaging.
\newblock {\em Optica}, 6(8):921--943, Aug 2019.

\bibitem{performance_DL}
Ruibo Shang, Kevin Hoffer-Hawlik, Fei Wang, Guohai Situ, and Geoffrey~P. Luke.
\newblock Two-step training deep learning framework for computational imaging
  without physics priors.
\newblock {\em Opt. Express}, 29(10):15239--15254, May 2021.

\bibitem{untrained_DL}
Kristina Monakhova, Vi~Tran, Grace Kuo, and Laura Waller.
\newblock Untrained networks for compressive lensless photography.
\newblock {\em Opt. Express}, 29(13):20913--20929, Jun 2021.

\bibitem{DIP_DL}
Victor Lempitsky, Andrea Vedaldi, and Dmitry Ulyanov.
\newblock Deep image prior.
\newblock In {\em 2018 IEEE/CVF Conference on Computer Vision and Pattern
  Recognition}, pages 9446--9454, 2018.

\bibitem{Y_Net}
Kaiqiang Wang, Jiazhen Dou, Qian Kemao, Jianglei Di, and Jianlin Zhao.
\newblock Y-net: a one-to-two deep learning framework for digital holographic
  reconstruction.
\newblock {\em Opt. Lett.}, 44(19):4765--4768, Oct 2019.

\bibitem{WPR}
Kaiqiang Wang, Li~Song, Chutian Wang, Zhenbo Ren, Guangyuan Zhao, Jiazhen Dou,
  Jianglei Di, George Barbastathis, Renjie Zhou, Jianlin Zhao, and Edmund~Y.
  Lam.
\newblock On the use of deep learning for phase recovery.
\newblock {\em Light: Science {\&} Applications}, 13(1):4, Jan 2024.

\bibitem{Non_Conv_DL}
Reinhard Heckel and Paul Hand.
\newblock Deep decoder: Concise image representations from untrained
  non-convolutional networks.
\newblock {\em International Conference on Learning Representations}, 2019.

\bibitem{Untrained_Blur}
Xiaole Tang, Xile Zhao, Jun Liu, Jianli Wang, Yuchun Miao, and Tieyong Zeng.
\newblock Uncertainty-aware unsupervised image deblurring with deep residual
  prior.
\newblock In {\em Proceedings of the IEEE/CVF Conference on Computer Vision and
  Pattern Recognition (CVPR)}, pages 9883--9892, June 2023.

\bibitem{SR_Untrained}
Wei Li, Ksenia Abrashitova, and Lyubov~V. Amitonova.
\newblock Super-resolution multimode fiber imaging with an untrained neural
  network.
\newblock {\em Opt. Lett.}, 48(13):3363--3366, Jul 2023.

\bibitem{Model_PR_AMP}
Li-Hao Yeh, Jonathan Dong, Jingshan Zhong, Lei Tian, Michael Chen, Gongguo
  Tang, Mahdi Soltanolkotabi, and Laura Waller.
\newblock Experimental robustness of {F}ourier ptychography phase retrieval
  algorithms.
\newblock {\em Opt. Express}, 23(26):33214--33240, Dec 2015.

\bibitem{GS_method}
G.-Z. Yang, B.-Z. Dong, B.-Y. Gu, J.-Y. Zhuang, and O.~K. Ersoy.
\newblock Gerchberg--{S}axton and {Y}ang--{G}u algorithms for phase retrieval
  in a nonunitary transform system: a comparison.
\newblock {\em Appl. Opt.}, 33(2):209--218, Jan 1994.

\bibitem{FB_method}
Timur~E. Gureyev, Timothy~J. Davis, Andrew Pogany, Sheridan~C. Mayo, and
  Stephen~W. Wilkins.
\newblock Optical phase retrieval by use of first {B}orn- and {R}ytov-type
  approximations.
\newblock {\em Appl. Opt.}, 43(12):2418--2430, Apr 2004.

\bibitem{Ill_posed_P}
Barbara Blaschke-Kaltenbacher and Heinz~W. Engl.
\newblock {\em Regularization Methods for Nonlinear Ill-Posed Problems with
  Applications to Phase Reconstruction}, pages 17--35.
\newblock Springer Vienna, Vienna, 1997.

\bibitem{ill_problem}
H~Egger and A~Leitão.
\newblock Nonlinear regularization methods for ill-posed problems with
  piecewise constant or strongly varying solutions.
\newblock {\em Inverse Problems}, 25(11):115014, oct 2009.

\bibitem{UNet_DL}
Olaf Ronneberger, Philipp Fischer, and Thomas Brox.
\newblock {U-Net}: Convolutional networks for biomedical image segmentation.
\newblock In Nassir Navab, Joachim Hornegger, William~M. Wells, and
  Alejandro~F. Frangi, editors, {\em Medical Image Computing and
  Computer-Assisted Intervention -- MICCAI 2015}, pages 234--241, Cham, 2015.
  Springer International Publishing.

\bibitem{3D_mesh}
C.~Osorio~Quero, D.~Durini, J.~Rangel-Magdaleno, J.~Martinez-Carranza, and
  R.~Ramos-Garcia.
\newblock Single-pixel near-infrared {3D} image reconstruction in outdoor
  conditions.
\newblock {\em Micromachines}, 13(5), 2022.

\bibitem{U2Net_DL}
Xuebin Qin, Zichen Zhang, Chenyang Huang, Masood Dehghan, Osmar Zaiane, and
  Martin Jagersand.
\newblock {$U^2$-Net}: Going deeper with nested u-structure for salient object
  detection.
\newblock {\em Pattern Recognition}, 106:107404, 2020.

\bibitem{Gdxray_dataset}
D.~Mery, V.~Riffo, U.~Zscherpel, G.~Mondragon, I.~Lillo, I.~Zuccar, H.~Lobel,
  and M.~Carrasco.
\newblock Gdxray: The database of {X}-ray images for nondestructive testing.
\newblock {\em Journal of Nondestructive Evaluation}, 34(42), 2015.

\bibitem{NR_IAQ}
Anish Mittal, Anush~Krishna Moorthy, and Alan~Conrad Bovik.
\newblock No-reference image quality assessment in the spatial domain.
\newblock {\em IEEE Transactions on Image Processing}, 21(12):4695--4708, 2012.

\bibitem{3D_RMSE}
R.R. Fulton, S.~Eberl, S.R. Meikle, B.F. Hutton, and M.~Braun.
\newblock A practical 3{D} tomographic method for correcting patient head
  motion in clinical spect.
\newblock {\em IEEE Transactions on Nuclear Science}, 46(3):667--672, 1999.

\bibitem{3D_metric}
Zicheng Zhang, Wei Sun, Xiongkuo Min, Tao Wang, Wei Lu, Wenhan Zhu, and
  Guangtao Zhai.
\newblock A no-reference visual quality metric for 3d color meshes.
\newblock In {\em 2021 IEEE International Conference on Multimedia \& Expo
  Workshops (ICMEW)}, pages 1--6, 2021.

\bibitem{PR_mathematic}
Jonathan Dong, Lorenzo Valzania, Antoine Maillard, Thanh-an Pham, Sylvain
  Gigan, and Michael Unser.
\newblock Phase retrieval: From computational imaging to machine learning: A
  tutorial.
\newblock {\em IEEE Signal Processing Magazine}, 40(1):45--57, 2023.

\bibitem{Fienup_PR}
J.~R. Fienup.
\newblock Phase retrieval algorithms: a comparison.
\newblock {\em Appl. Opt.}, 21(15):2758--2769, Aug 1982.

\bibitem{CH_PR_F}
Kejun Huang, Yonina~C. Eldar, and Nicholas~D. Sidiropoulos.
\newblock Phase retrieval from 1{D} {F}ourier measurements: Convexity,
  uniqueness, and algorithms.
\newblock {\em IEEE Transactions on Signal Processing}, 64(23):6105--6117,
  2016.

\bibitem{Yiel_PR_F}
Wenxiang Cong, Nanxian Chen, and Benyuan Gu.
\newblock A recursive method for phase retrieval in {F}ourier transform domain.
\newblock {\em Chinese Science Bulletin}, 43(1):40--44, Aug 1998.

\bibitem{UNNphase}
Gauri Jagatap and Chinmay Hegde.
\newblock Phase retrieval using untrained neural network priors.
\newblock In {\em NeurIPS 2019 Workshop on Solving Inverse Problems with Deep
  Networks}, 2019.

\bibitem{WF_PR}
Emmanuel~J. Candès, Xiaodong Li, and Mahdi Soltanolkotabi.
\newblock Phase retrieval via {W}irtinger flow: Theory and algorithms.
\newblock {\em IEEE Transactions on Information Theory}, 61(4):1985--2007,
  2015.

\bibitem{AMP_PR}
Junjie Ma, Ji~Xu, and Arian Maleki.
\newblock Optimization-based amp for phase retrieval: The impact of
  initialization and $\ell_{2}$ regularization.
\newblock {\em IEEE Transactions on Information Theory}, 65(6):3600--3629,
  2019.

\bibitem{Fourier_PR}
M.~Hayes.
\newblock The reconstruction of a multidimensional sequence from the phase or
  magnitude of its {F}ourier transform.
\newblock {\em IEEE Transactions on Acoustics, Speech, and Signal Processing},
  30(2):140--154, 1982.

\bibitem{DL_PR_Methods}
Daniele Orsuti, Cristian Antonelli, Alessandro Chiuso, Marco Santagiustina,
  Antonio Mecozzi, Andrea Galtarossa, and Luca Palmieri.
\newblock Deep learning-based phase retrieval scheme for minimum-phase signal
  recovery.
\newblock {\em Journal of Lightwave Technology}, 41(2):578--592, 2023.

\bibitem{DeepPhaseCut}
Eunju Cha, Chanseok Lee, Mooseok Jang, and Jong~Chul Ye.
\newblock {DeepPhaseCut}: Deep relaxation in phase for unsupervised fourier
  phase retrieval.
\newblock {\em IEEE Transactions on Pattern Analysis and Machine Intelligence},
  44(12):9931--9943, 2022.

\bibitem{Oversampled_Fourier}
Longlong Wu, Shinjae Yoo, Ana~F. Suzana, Tadesse~A. Assefa, Jiecheng Diao,
  Ross~J. Harder, Wonsuk Cha, and Ian~K. Robinson.
\newblock Three-dimensional coherent {X}-ray diffraction imaging via deep
  convolutional neural networks.
\newblock {\em npj Computational Materials}, 7(1):175, Oct 2021.

\bibitem{GAN_PR}
Yuhe Zhang, Mike~Andreas Noack, Patrik Vagovic, Kamel Fezzaa, Francisco
  Garcia-Moreno, Tobias Ritschel, and Pablo Villanueva-Perez.
\newblock Phase{GAN}: a deep-learning phase-retrieval approach for unpaired
  datasets.
\newblock {\em Opt. Express}, 29(13):19593--19604, Jun 2021.

\bibitem{ResNet_FR}
Kaiming He, Xiangyu Zhang, Shaoqing Ren, and Jian Sun.
\newblock Deep residual learning for image recognition.
\newblock In {\em 2016 IEEE Conference on Computer Vision and Pattern
  Recognition (CVPR)}, pages 770--778, 2016.

\bibitem{Lensless_FR}
Yukuan Yang, Lei Deng, Peng Jiao, Yansong Chua, Jing Pei, Cheng Ma, and Guoqi
  Li.
\newblock Transfer learning in general lensless imaging through scattering
  media.
\newblock In {\em 2020 15th IEEE Conference on Industrial Electronics and
  Applications (ICIEA)}, pages 1132--1141, 2020.

\bibitem{FR_DL_2}
Chang-Jen Wang, Chao-Kai Wen, Shang-Ho Tsai, Shi Jin, and Geoffrey~Ye Li.
\newblock Phase retrieval using expectation consistent signal recovery
  algorithm based on hypernetwork.
\newblock {\em IEEE Transactions on Signal Processing}, 69:5770--5783, 2021.

\bibitem{LSpetral}
David Morales, Andr\'{e}s Jerez, and Henry Arguello.
\newblock Learning spectral initialization for phase retrieval via deep neural
  networks.
\newblock {\em Appl. Opt.}, 61(9):F25--F33, Mar 2022.

\bibitem{Book_FR_DL}
Tobias Uelwer, Tobias Hoffmann, and Stefan Harmeling.
\newblock Non-iterative phase retrieval with cascaded neural networks.
\newblock In Igor Farka{\v{s}}, Paolo Masulli, Sebastian Otte, and Stefan
  Wermter, editors, {\em Artificial Neural Networks and Machine Learning --
  ICANN 2021}, pages 295--306, Cham, 2021. Springer International Publishing.

\bibitem{FR_Ite_DL}
\c{C}a\u{g}atay I\c{s}{i}l, Figen~S. Oktem, and Aykut Ko\c{c}.
\newblock Deep iterative reconstruction for phase retrieval.
\newblock {\em Appl. Opt.}, 58(20):5422--5431, Jul 2019.

\bibitem{Book_convolutional}
Wangmeng Zuo, Kai Zhang, and Lei Zhang.
\newblock {\em Convolutional Neural Networks for Image Denoising and
  Restoration}, pages 93--123.
\newblock Springer International Publishing, Cham, 2018.

\bibitem{Unet_PR_F1}
Feixiang Luo, Jun Wang, Jie Zeng, Lu~Zhang, Boyu Zhang, Kuiwen Xu, and Xiling
  Luo.
\newblock Cascaded complex {U-Net} model to solve inverse scattering problems
  with phaseless-data in the complex domain.
\newblock {\em IEEE Transactions on Antennas and Propagation},
  70(8):6160--6170, 2022.

\bibitem{Unet_PR_F2}
Teng Zhang.
\newblock Phase retrieval by alternating minimization with random
  initialization.
\newblock {\em IEEE Transactions on Information Theory}, 66(7):4563--4573,
  2020.

\bibitem{Unet_PR_F3}
Feilong Zhang, Xianming Liu, Cheng Guo, Shiyi Lin, Junjun Jiang, and Xiangyang
  Ji.
\newblock Physics-based iterative projection complex neural network for phase
  retrieval in lensless microscopy imaging.
\newblock In {\em 2021 IEEE/CVF Conference on Computer Vision and Pattern
  Recognition (CVPR)}, pages 10518--10526, 2021.

\bibitem{Unet_PR}
Shasha Pu, Lan Li, Yu~Xiang, and Xiaolong Qiu.
\newblock Phase retrieval based on enhanced generator conditional generative
  adversarial network.
\newblock In {\em 2022 4th International Conference on Intelligent Control,
  Measurement and Signal Processing (ICMSP)}, pages 825--829, 2022.

\bibitem{Unet_PR_F4}
Yoshiki Masuyama, Kohei Yatabe, Yuma Koizumi, Yasuhiro Oikawa, and Noboru
  Harada.
\newblock Deep {Griffin–Lim} iteration: Trainable iterative phase
  reconstruction using neural network.
\newblock {\em IEEE Journal of Selected Topics in Signal Processing},
  15(1):37--50, 2021.

\bibitem{Unet_PR_F5}
Shasha Pu, Lan Li, Yu~Xiang, and Xiaolong Qiu.
\newblock Phase retrieval based on enhanced generator conditional generative
  adversarial network.
\newblock In {\em 2022 4th International Conference on Intelligent Control,
  Measurement and Signal Processing (ICMSP)}, pages 825--829, 2022.

\bibitem{Unet_vanish}
Peizhi Wen, Menglong Sun, and Yongqing Lei.
\newblock An improved {U-Net} method for sequence images segmentation.
\newblock In {\em 2019 Eleventh International Conference on Advanced
  Computational Intelligence (ICACI)}, pages 184--189, 2019.

\bibitem{eikonal}
Silvia Tozza and Maurizio Falcone.
\newblock Analysis and approximation of some shape-from-shading models for
  non-{L}ambertian surfaces.
\newblock {\em Journal of Mathematical Imaging and Vision}, 55(2):153--178, Jun
  2016.

\bibitem{Lax_3D}
Chiu~Yen Kao, Stanley Osher, and Jianliang Qian.
\newblock Lax–{F}riedrichs sweeping scheme for static {H}amilton–{J}acobi
  equations.
\newblock {\em Journal of Computational Physics}, 196(1):367--391, 2004.

\bibitem{UNET_keras}
Nahian Siddique, Sidike Paheding, Colin~P. Elkin, and Vijay Devabhaktuni.
\newblock {U-Net} and its variants for medical image segmentation: A review of
  theory and applications.
\newblock {\em IEEE Access}, 9:82031--82057, 2021.

\bibitem{Brisque_NR_IAQ}
Anish Mittal, Anush~K. Moorthy, and Alan~C. Bovik.
\newblock Blind/referenceless image spatial quality evaluator.
\newblock In {\em 2011 Conference Record of the Forty Fifth Asilomar Conference
  on Signals, Systems and Computers (ASILOMAR)}, pages 723--727, 2011.

\bibitem{NIQE_NR_IAQ}
Anish Mittal, Rajiv Soundararajan, and Alan~C. Bovik.
\newblock Making a “completely blind” image quality analyzer.
\newblock {\em IEEE Signal Processing Letters}, 20(3):209--212, 2013.

\bibitem{3D_FOM}
Chris~J. Budd, Andrew~T.T. McRae, and Colin~J. Cotter.
\newblock The scaling and skewness of optimally transported meshes on the
  sphere.
\newblock {\em Journal of Computational Physics}, 375:540--564, 2018.

\bibitem{GAN_DL}
Yuhe Zhang, Mike~Andreas Noack, Patrik Vagovic, Kamel Fezzaa, Francisco
  Garcia-Moreno, Tobias Ritschel, and Pablo Villanueva-Perez.
\newblock Phase{GAN}: a deep-learning phase-retrieval approach for unpaired
  datasets.
\newblock {\em Opt. Express}, 29(13):19593--19604, Jun 2021.

\end{thebibliography}

%%%%%%%%%% If preparing manually:
% \begin{thebibliography}{1}
	% \newcommand{\enquote}[1]{``#1''}
	
	% \bibitem{Zhang:14}
	% Y.~Zhang, S.~Qiao, L.~Sun, Q.~W. Shi, W.~Huang, L.~Li, and Z.~Yang,
	%   \enquote{Photoinduced active terahertz metamaterials with nanostructured
		%   vanadium dioxide film deposited by sol-gel method,}
	%   {\protect\JournalTitle{Optics Express}} \textbf{22}, 11070--11078 (2014).
	
	% \bibitem{Optica}
	% {Optica}, \enquote{{Optica Publishing Group},}
	%   \url{http://www.opg.optica.org}.
	
	% \bibitem{FORSTER2007}
	% P.~Forster, V.~Ramaswamy, P.~Artaxo, T.~Bernsten, R.~Betts, D.~Fahey,
	%   J.~Haywood, J.~Lean, D.~Lowe, G.~Myhre, J.~Nganga, R.~Prinn, G.~Raga,
	%   M.~Schulz, and R.~V. Dorland, \enquote{Changes in atmospheric consituents and
		%   in radiative forcing,} in \enquote{Climate Change 2007: The Physical Science
		%   Basis. Contribution of Working Group 1 to the Fourth assesment report of
		%   Intergovernmental Panel on Climate Change,}  S.~Solomon, D.~Qin, M.~Manning,
	%   Z.~Chen, M.~Marquis, K.~B. Averyt, M.~Tignor, and H.~L. Miler, eds.
	%   (Cambridge University Press, 2007).
	
	% \end{thebibliography}

\end{document}